\documentclass[fleqn,usenatbib]{mnras}
\usepackage[T1]{fontenc}
\usepackage{ae,aecompl}
\usepackage{graphicx}	
\usepackage{amsmath}	
\usepackage{amssymb}	
\usepackage{multirow}
\usepackage{siunitx}
\usepackage{txfontsb} 
\usepackage{upgreek}
\usepackage{color} 
\usepackage{glossaries}
\glsdisablehyper

\newcommand{\ud}{\rmn{d}} 
\newcommand{\lcdm}{\gls{lcdm}}
\DeclareSIUnit\parsec{pc}
\title[J-PAS forecasts on interacting dark energy]{J-PAS: forecasts on interacting dark energy from baryon acoustic oscillations and redshift-space distortions}
\author[A. A. Costa et al.]{
	A. A. Costa$^{1,2}$\thanks{E-mail: alencar@if.usp.br},
    R. J. F. Marcondes$^{1}$,
	R. G. Landim$^{1,3}$,
    E. Abdalla$^1$,
    L. R. Abramo$^1$,
    \newauthor
    H. S. Xavier$^4$,
    A. A. Orsi$^5$,
    N. Chandrachani Devi$^6$,
    A. J. Cenarro$^7$,
    D. Crist\'obal-Hornillos$^7$,
    \newauthor
    R. A. Dupke$^{8,9}$,
    A. Ederoclite$^4$,
    A. Mar\'in-Franch$^7$,
    C. M. Oliveira$^4$,
    H. V\'azquez Rami\'o$^5$,
    \newauthor
    K. Taylor$^{10}$ and
    J. Varela$^7$
    \\
$^{1}$Instituto de F\'{i}sica, Universidade de S\~{a}o Paulo, Rua do Mat\~ao 1371, S\~{a}o Paulo, SP 05508-090, Brazil \\
$^{2}$Center for Gravitation and Cosmology, College of Physical Science and Technology, Yangzhou University, Yangzhou 225009, China \\
$^{3}$SLAC National Accelerator Laboratory, 2575 Sand Hill Rd., Menlo Park, CA 94025 USA \\
$^{4}$Instituto de Astronomia, Geof\'{i}sica e Ci\^{e}ncias Atmosf\'{e}ricas, Universidade de S\~{a}o Paulo, Rua do Mat\~ao 1226, S\~{a}o Paulo, SP 05508-090, Brazil \\
$^{5}$Centro de Estudios de F\'{i}sica del Cosmos de Arag\'{o}n, Plaza de San Juan 1, E-44001 Teruel, Spain \\
$^{6}$Instituto de Astronom\'ia, Universidad Nacional Aut\'onoma de M\'exico, A. P. 70-264, 04510, M\'exico, D.F., M\'exico \\
$^{7}$Centro de Estudios de F\'{i}sica del Cosmos de Arag\'{o}n (CEFCA), Unidad Asociada al CSIC, Plaza de San Juan 1, E-44001 Teruel, Spain \\
$^{8}$Observatorio Nacional-MCTIC, Rua Jose Cristino 77, Rio de Janeiro, Brazil \\
$^{9}$Dept. of Astronomy, University of Michigan, 1085 S. University, Ann Arbor \\
$^{10}$Instruments4, 4121 Pembury Place, La Ca\~nada-Flintridge, Ca 91011, USA \\
}
\date{Accepted 2019 June 12. Received 2019 June 12; in original form  2019 April 2.}
\pubyear{2019}

\newacronym[shortplural=BAO]{bao}{BAO}{baryon acoustic oscillation}
\newacronym{lrg}{LRG}{luminous red galaxies}
\newacronym{elg}{ELG}{emission line galaxies}
\newacronym{qso}{QSO}{quasars}
\newacronym{jpas}{J-PAS}{Javalambre-Physics of the Accelerated Universe Astrophysical Survey}
\newacronym{lcdm}{$\Lambda\rmn{CDM}$}{$\Lambda$-Cold Dark Matter}
\newacronym{de}{DE}{dark energy}
\newacronym{dm}{DM}{dark matter}
\newacronym{flrw}{FLRW}{Friedmann--Lema\^itre--Robertson--Walker}
\newacronym[shortplural=RSD]{rsd}{RSD}{redshift-space distortion}
\newacronym{eos}{EoS}{equation of state}
\newacronym{sdss}{SDSS}{Sloan Digital Sky Survey}

\usepackage{etoolbox}
\makeatletter
\patchcmd\@combinedblfloats{\box\@outputbox}{\unvbox\@outputbox}{}{%
   \errmessage{\noexpand\@combinedblfloats could not be patched}%
}%
 \makeatother

\begin{document}

\label{firstpage}
\pagerange{\pageref{firstpage}--\pageref{lastpage}}
\maketitle

\begin{abstract}
We estimate the constraining power of J-PAS for
parameters of an interacting dark energy cosmology. The
survey is expected to map several millions of luminous red
galaxies, emission line galaxies and quasars in an area of
thousands of square degrees in the northern sky with precise
photometric redshift measurements. 
Forecasts for the DESI and Euclid surveys are also
evaluated and compared to J-PAS. With the Fisher matrix
approach, we find that J-PAS can place constraints on the
interaction parameter comparable to those from DESI, with an
absolute uncertainty of about $0.02$, when the interaction
term is proportional to the dark matter energy density, and
almost as good, of about $0.01$, when the interaction is
proportional to the dark energy density. 
For the equation of state of dark energy, the constraints from J-PAS are slightly better
in the two cases (uncertainties $0.04$--$0.05$ against
$0.05$--$0.07$ around the fiducial value $-1$).
Both surveys stay behind Euclid but follow it closely, imposing
comparable constraints in all specific cases considered.
\end{abstract}


\section{Introduction}
\label{Sec:intro}
The lack of knowledge regarding the nature of the dark sector, especially the
cosmic acceleration \citep{riess_observational_1998,
    perlmutter_measurements_1999}, has led to a continuous endeavor to understand
the origin of such accelerated expansion and its dynamics. Several ongoing and upcoming spectroscopic,
photometric and radio surveys have been proposed to address this problem, including DES
\citep{the_dark_energy_survey_collaboration_dark_2005},
LSST \citep{lsst_science_collaboration_lsst_2009},
eBOSS \citep{dawson_sdss-iv_2016},
DESI \citep{desi_collaboration_desi_2016},
Euclid \citep{laureijs_euclid_2011},
BINGO \citep{battye_bingo:_2012, wuensche_bingo_2018} and
SKA \citep{maartens_overview_2015}. Among them, the \glsentrylong{jpas} 
\citep[\glsentryshort{jpas}\glsunset{jpas},][]{Benitez:2008fs,benitez_j-pas:_2014}
is a multi narrow-band photometric survey which
will cover up to \num{8500} square degrees of the northern
sky and measure $0.003 \left(1+z \right)$ precision photometric
redshifts for $9\times10^7$ \gls{lrg} and \gls{elg} plus
several millions of \gls{qso}.
In addition, it aims to detect and measure the mass of $7 \times 10^5$ galaxy
clusters and groups, improving the constrains on dark energy.

On the theoretical side, deviations from the \lcdm\ model have been
proposed over the years, whose alternatives to the cosmological constant include
canonical and non-canonical scalar fields
\citep{peebles_cosmology_1988, ratra_cosmological_1988, frieman_late-time_1992,frieman_cosmology_1995, caldwell_cosmological_1998}, holographic dark energy \citep{hsu_entropy_2004, li_model_2004}, vector fields \citep{ArmendarizPicon:2004pm}, metastable dark energy \citep{stojkovic_dark_2008}, among others. One interesting possibility to consider is when we allow an exchange of energy-momentum between the two components of the dark sector \citep{wetterich_asymptotically_1995, amendola_coupled_2000}.
This mechanism could be one reason why \gls{de} and \gls{dm} contribute to the
present Universe with comparable energy densities, alleviating the coincidence problem \citep{zimdahl_interacting_2001, chimento_interacting_2003}. Models of interacting \gls{de} have been widely explored in the literature \citep{wang_dark_2016}.

In this work, we consider a phenomenological description of the
\gls{de}-\gls{dm} interaction and use the Fisher matrix formalism to assess the capability of baryon acoustic oscillations (BAO) and redshift-space distortions (RSD), as observed by J-PAS, to improve the
constraints on the \gls{eos} of dark energy and on the coupling constant.
We also take advantage of the added power from combining
multiple tracers of large-scale structure in order to
improve the accuracy of measurements of the matter growth
rate \citep{abramo_why_2013, abramo_fourier_2016,
    marin_bosswigglez_2016, witzemann_simulated_2018}. Our results are compared with those obtained for the Euclid and DESI surveys using the same methodology.

The paper is organized as follows. Section~\ref{sec:model} introduces the interacting model, which we analyse in three specific cases. In Section~\ref{sec:data}, we describe the details of the surveys considered here. Section~\ref{sec:RSD} explains how the \gls{rsd} parameter changes in a interacting \gls{de} model. Our Fisher matrix analysis is presented in section~\ref{sec:Fisher} and section~\ref{sec:results} contains our results, including a comparison with forecasts from the DESI and Euclid surveys. Section~\ref{sec:conclusions} is reserved for conclusions.

\section{The interacting dark energy model}
\label{sec:model}
The dark sector constitutes about 95 per cent of the energy density of the
Universe. Its components, dubbed dark matter and dark energy, do not have a
definitive model yet. Besides, their energy densities are of the same order of
magnitude despite the fact they evolve completely different in the standard
\lcdm\ model. Hence, it seems natural to assume they can interact with each
other. In this case, the energy-momentum tensors $T^{\mu\nu}_{(\lambda)}$
of each component $\lambda$ are not independently conserved anymore,
\begin{align}
	\label{eq:T}
	\nabla_\mu T^{\mu\nu}_{(\lambda)} = Q^\nu_{(\lambda)},
\end{align}
where $Q^\nu_{(\lambda)}$ is the four-vector that accounts for the coupling and
satisfies the constraint $\sum_\lambda Q^\nu_{(\lambda)} = 0$.

Assuming a flat Friedmann-Lemaitre-Robertson-Walker (FLRW) universe, the
conservation equations (\ref{eq:T}) give rise to the continuity equations
\citep{marcondes_analytic_2016, costa_constraints_2017} 
\begin{gather}
	\label{eq:continuity}
    \begin{aligned}
    	\dot\rho_{c} &+3H\rho_{c} = Q, \\
		\dot\rho_d &+3H \left(1+w \right)\rho_d = -Q,
    \end{aligned}
\end{gather}
where $H=\dot{a}/a$ is the Hubble rate, $Q$ is the coupling and $\rho_c$ and
$\rho_d$ are the background energy densities of \gls{dm} and \gls{de}, respectively.
The \gls{de} \gls{eos} is given by $w = P_d/\rho_d$, where $P_d$ is its
pressure. Throughout this work a dot represents derivative with respect to the cosmic
time $t$.

From the continuity equations (\ref{eq:continuity}), we see that a positive $Q$
indicates an energy transfer from \gls{de} to \gls{dm}. General interactions
including a field derived from Lagrangian models have been considered in other
works \citep{micheletti_field_2009, costa_quintessence_2015,damico_quantum_2016,
    landim_metastable_2017}. However, as we still do not know the correct theory
to describe \gls{dm} and \gls{de}, we can investigate an interaction between
them from phenomenological arguments. In this work, we assume a phenomenological
coupling $Q$ which, in the generic case, have contributions proportional to the
\gls{dm} and \gls{de} densities
\begin{align}
\label{eq:coupling} 
Q = 3H \left(\xi_c\rho_c + \xi_d\rho_d \right),
\end{align}
where $\xi_c$ and $\xi_d$ are the corresponding coupling constants.

Interacting \gls{de} models with constant \gls{eos} have already been shown to
suffer from instabilities with respect to curvature and \gls{de} perturbations
\citep{valiviita_large-scale_2008, he_stability_2009}. Table~\ref{tab:stabilities} summarizes the
allowed regions for the interaction and the \gls{de} \gls{eos}
parameters as shown by \citet{he_stability_2009} and \citet{gavela_dark_2009}.
However, this is likely a problem related to the oversimplicity of the
interaction, which can be overcome in a more sophisticated Lagrangian description
as in \citet{costa_quintessence_2015}. In fact, \cite{Yang:2018euj} yield a phenomenological model with an interaction dependent on the DE equation of state, which is stable in the whole parameter space with an interaction parameter greater than zero. See also \citet{wang_dark_2016} for a review on
interacting models.
\begin{table}
    \setlength\tabcolsep{16pt} 
    \centering 
    \caption{Stability conditions on the \gls{eos} and interaction sign for the
        phenomenological interacting \gls{de} model.}
     \label{tab:stabilities}    
    \renewcommand*{\arraystretch}{1.4}
    \begin{tabular}{@{}lc@{}}
        \hline 
        Case     &   Constant \gls{eos} and  interaction sign   \\ 
        \hline
        \multirow{2}{*}{$Q \propto \rho_{d}$ ($\xi_c = 0$)} & $w < -1$ and   $\xi_d > 0$ ; or \\ 
        & $-1 < w < 0$ and $\xi_d < 0$ \\
        \hline 
        \multirow{2}{*}{$Q \propto \rho_{c}$ ($\xi_d = 0$)}  & \multirow{2}{*}{$w < -1, \, \forall \, \xi_c $}       \\
        & \\
        \hline 
        \multirow{2}{*}{$\xi_d \ne 0$ and $\xi_c \ne 0$}  &   \multirow{2}{*}{$w < -1, \, \xi_d > 0, \, \forall \, \xi_c$}       \\ 
        &   \\ 
        \hline
    \end{tabular}
\end{table}

\section{The Data Set}
\label{sec:data}
The data considered correspond to the two-point function or, more precisely, the
power spectrum of the clustering of some type of galaxy or quasar. \gls{jpas} will
be able to detect millions of \glsentrylong{lrg}, \glsentrylong{elg} and
\glsentrylong{qso}. Table~\ref{tab:tracer_densities} gives the expected number densities as
a function of redshift for different tracers. In the plane-parallel (distant
observer, $k^2 = k_{\parallel}^2 + k_{\bot}^2$) approximation, the observed
galaxy power spectrum is given by \citep{seo_probing_2003, wang_designing_2010}
\begin{align}
\label{eq:powerspectrum} 
P_{g,\,\rmn{obs}} = \left[\frac{D_A^{\rmn{fid}}(z)}{D_A(z)} \right]^2\frac{H(z)}{H^{\rmn{fid}}(z)} \left[ \sigma_{8,\,g}(z) + b \, \beta(z) \, \sigma_{8,\,m}\, \mu^2 \right]^2 C(k) + P_{\rmn{shot}}.
\end{align}
The prefactors, due to the Alcock-Paczynski effect
\citep{alcock_evolution_1979}, account for deviations from the fiducial Hubble
rate and angular diameter distance to the true cosmology ones. $\sigma_{8,\, g} = b \, \sigma_{8,\, m}$, where $\sigma_{8,\, m}$ is the variance of the matter density field averaged in spheres of radius $8~h^{-1}\si{Mpc}$ and $b$ is a bias between matter and galaxy overdensities. The RSD parameter $\beta$ is equal to the matter growth rate divided by the bias, $f_m/b$. $\mu =
k_{\parallel} / k$ is the cosine of the angle between the
wavevector and the line of sight, $C(k) \equiv P_m(k,z)/\sigma_{8,\,m}^2(z) =
P_{0,\,m}(k)/\sigma_{8,\,0}^2$ is the normalized true matter power spectrum and
$P_{\rmn{shot}}$ parametrizes a residual shot noise.
\begin{table}
    \setlength\tabcolsep{16pt}
    \centering
    \caption{Number densities of \glsentrylong{lrg}, \glsentrylong{elg} and
        \glsentrylong{qso} for \gls{jpas}, in units of $10^{-5} \, h^{3}$\,\si{\per\cubic\mega\parsec}. The factor $10^{-5}$
        in the unit here and in the following tables is to allow a better
        comparison.}
    \label{tab:tracer_densities}    
    \sisetup{
        round-mode=places,
        round-precision=1,
        table-format=1.1
    }
    \begin{tabular}{@{}SS[table-omit-exponent,fixed-exponent=-5]S[table-omit-exponent,fixed-exponent=-5]S[table-omit-exponent,fixed-exponent=-5,round-precision=2,table-format=1.2]@{}}
        \hline
        {$z$} &   {\gls{lrg}} &   {\gls{elg}}   &   {\gls{qso}} \\
        \hline
        0.3 &   0.00226575   &   0.0295855  &  4.536446268875312e-6 \\
        0.5 &   0.00156325  &   0.0118113  & 1.1395037830060143e-5 \\
        0.7 &   0.000688     &   0.00502125   & 1.609591875735657e-5 \\
        0.9 &   0.00011975   &   0.0013795  &  2.27360895599638e-5 \\
        1.1 &   0.000009      &   0.000412     & 2.8623040911866468e-5 \\
        1.3 &   0         &   0.00006725   & 3.603427356677276e-5 \\
        1.5 &   0         &   0      &  3.603427356677276e-5 \\
        1.7 &   0         &   0         & 3.211558012135426e-5 \\
        1.9 &   0         &   0         & 2.8623040911866468e-5 \\
        2.1 &   0         &   0         & 2.551031206494158e-5 \\
        2.3 &   0         &   0         & 2.27360895599638e-5 \\
        2.5 &   0         &   0         & 2.026356114981056e-5 \\
        2.7 &   0         &   0         & 1.8059917884699187e-5 \\
        2.9 &   0         &   0         & 1.609591875735657e-5 \\
        3.1 &   0         &   0         & 1.4345502692618598e-5 \\
        3.3 &   0         &   0         & 1.2785442732796502e-5 \\
        3.5 &   0         &   0         & 1.1395037830060143e-5 \\
        3.7 &   0         &   0         & 9.051400284168088e-6 \\
        3.9 &   0         &   0         & 7.189782809506104e-6  \\
        \hline
    \end{tabular}
\end{table}

The galaxy overdensity is related to the matter overdensity through a bias,
$\delta_g = b(k,z) \,\delta_m$, which in general can be a function of the scale
and redshift. Here, we assume that the bias
    depends only on the redshift and is given, for each
    tracer, by \citep{ross_clustering_2009,
        desi_collaboration_desi_2016}
\begin{gather}
    \label{eq:biases}
    \begin{aligned}
        b_{\rmn{LRG}}(z) &= \frac{1.7}{D(z)}, \quad 
        b_{\rmn{ELG}}(z) = \frac{0.84}{D(z)}, \\
        b_{\rmn{QSO}}(z) &= 0.53 + 0.289 \left(1 + z\right)^2,
    \end{aligned}
\end{gather}
where $D(z) = \exp \left[-\int_0^z \ud z' \, f_m/\left(1 + z'\right)\right]$
    is the growth factor normalized to 1 today and $f_m = \ud \ln{\delta_m}/\ud\ln{a}$ is the matter growth rate. In our calculations we use weak Gaussian priors for the biases, with variances $\sigma_b = \num{0.5}$.

We also compare our results for J-PAS with the expected results from DESI and Euclid.
The number densities we are assuming are presented in
Table~\ref{tab:tracers_DESI} for the DESI survey and in
Table~\ref{tab:tracers_Euclid} for Euclid.
DESI has the same bias as those in eq. (\ref{eq:biases}); on the other
hand, we use $b(z) = \sqrt{1+z}$ in the case of Euclid \citep{laureijs_euclid_2011, wang_designing_2010,orsi_probing_2010}. This choice of bias is a good approximation to studies from semianalytic models of galaxy formation as in \citet{orsi_probing_2010} \citep[see, for example,][]{giannantonio_constraining_2012}. Although this choice is different from the one made for J-PAS and DESI, our results are only weakly dependent on it as we are considering information from the BAO wiggles only \citep{rassat_deconstructing_2008}.
\begin{table}
    \setlength\tabcolsep{16pt}
    \centering
    \caption{Number densities of \glsentrylong{lrg}, \glsentrylong{elg} and
        \glsentrylong{qso} for DESI, in units of
        $10^{-5} \, h^{3}$\,\si{\per\cubic\mega\parsec}.}
    \label{tab:tracers_DESI}    
    \sisetup{
        round-mode=figures,
        round-precision=2,
        table-format=1.1
    }
    \begin{tabular}{@{}S[round-mode=places,table-format=1.2]S[table-omit-exponent,fixed-exponent=-5]S[table-omit-exponent,fixed-exponent=-5]S[table-omit-exponent,fixed-exponent=-5]@{}}
        \hline
        {$z$} & {\gls{lrg}} & {\gls{elg}}  & {\gls{qso}} \\
        \hline
         0.65 & 4.9e-4 & 1.8e-4 & 2.8e-5 \\ 
         0.75 & 4.9e-4 & 1.12e-3 & 2.7e-5 \\ 
         0.85 & 2.9e-4 & 8.3e-4 & 2.6e-5 \\ 
         0.95 & 1.0e-4 & 8.1e-4 & 2.6e-5 \\ 
         1.05 & 2.e-5 & 5.1e-4 & 2.6e-5 \\ 
         1.15 & 1.e-5 & 4.5e-4 & 2.5e-5 \\ 
         1.25 & 0    & 4.2e-4 & 2.5e-5 \\ 
         1.35 & 0    & 1.5e-4 & 2.5e-5 \\
         1.45 & 0    & 1.3e-4 & 2.4e-5 \\
         1.55 & 0    & 9.e-5 & 2.4e-5 \\
         1.65 & 0    & 3.e-5 & 2.3e-5 \\
         1.75 & 0    & 0    & 2.3e-5 \\
         1.85 & 0   & 0    & 2.2e-5 \\
         \hline
    \end{tabular}
\end{table}
\begin{table}
    \setlength\tabcolsep{16pt}
    \centering
    \caption{Number densities of \glsentrylong{elg} for Euclid, in units of
        $10^{-5} \, h^{3}$\,\si{\per\cubic\mega\parsec}.}
    \label{tab:tracers_Euclid}    
        \sisetup{
        round-mode=places,
        round-precision=3,
        table-format=1.2
    }
    \begin{tabular}{@{}S[round-mode=places,round-precision=1,table-format=1.1]S[table-omit-exponent,fixed-exponent = -5]@{}}
        \hline
        {$z$} & {\gls{elg}} \\
        \hline 
        0.6 & 3.56e-3 \\ 
        0.8 & 2.42e-3 \\ 
        1.0 & 1.81e-3 \\
        1.2 & 1.44e-3 \\
        1.4 & 0.99e-3 \\ 
        1.8 & 0.33e-3 \\ 
        \hline
    \end{tabular}
\end{table}

Two scenarios are considered for the J-PAS survey, a more conservative initial expectation with a survey area of \SI{4000}{\square\deg} and a possible future best case scenario with \SI{8500}{\square\deg}. The DESI and Euclid survey areas are estimated as \SIlist{14000;15000}{\square\deg}, respectively.
The redshift errors are assumed to be $0.003 \left(1+z \right)$ for \gls{jpas} and $0.001 \left(1+z \right)$ for DESI and Euclid.

\section{The modified RSD parameter}
\label{sec:RSD}
Measurements of $\beta$ from the power spectra or
peculiar velocities are based on its correspondence with the velocity divergence
$\theta$ as established by the continuity equation. Since this equation is
violated in interacting models, we must make sure to use the correct quantity
that corresponds to the velocity field when confronting our model with
observations or making forecasts for some experiment \citep[see, for
example,][]{marcondes_analytic_2016, borges_growth_2017, kimura_are_2018}.

For an interacting \gls{de} model with coupling given by
eq.~(\ref{eq:coupling}), the continuity equation for \gls{dm} at first order in perturbations in the sub-horizon limit ($k \gg H$) reads
\begin{align}
    \delta'_{c} + 3 \mathcal{H} \xi_d \frac{\rho_d}{\rho_c} \left(\delta_c - \delta_d \right) + \theta_c = 0 .
\end{align}
In this equation, we now express, for convenience, the evolution in terms of the conformal time $\tau$, with the prime representing $\ud/\ud \tau$ and $\mathcal{H} = a'/a$. The total matter density is $\rho_m = \rho_b + \rho_c$ and its perturbation $\delta_m = \left(\rho_b \delta_b + \rho_c \delta_c \right)/ \rho_m$. Thus, its (conformal) time derivative is given by
\begin{align} 
    \rho_m \delta_m' = - 3 \mathcal{H} \xi_c \rho_c \left( \delta_m - \delta_c \right) - 3 \mathcal{H} \xi_d \rho_d \left( \delta_m - \delta_d \right) - (\rho_b \theta_b + \rho_c \theta_c) ,
\end{align}
where the continuity equations for baryons and \gls{dm} have been used. This expression can be rewritten as
\begin{align} 
    \mathcal{H} &\left[ \frac{\ud \ln \delta_m}{\ud \ln a} + 3 \xi_c \frac{ \rho_c}{\rho_m} \left(1 - \frac{\delta_c}{\delta_m} \right) + 3 \xi_d \frac{\rho_d}{\rho_m} \left( 1 - \frac{\delta_d}{\delta_m} \right) \right] \delta_m + {} \nonumber \\
    {} &+ \frac{\rho_b \theta_b + \rho_c \theta_c}{\rho_m} = 0.
\end{align}

We can now recognize the term $\left(\rho_b \theta_b + \rho_c \theta_c \right)/\rho_m$ as $\theta_m$, as usual, and express the continuity equation corrected for the interaction
\begin{align}
    \mathcal{H} \tilde{f}_m \delta_m + \theta_m = 0 ,
\end{align} 
where 
\begin{align} 
    \tilde{f}_m \equiv \frac{\ud \ln \delta_m}{\ud \ln a} + 3 \left( \frac{\xi_c \rho_c + \xi_d \rho_d}{\rho_m} - \frac{\xi_c \rho_c \delta_c + \xi_d \rho_d \delta_d}{\rho_m \delta_m}\right)
\end{align}
is the growth rate for the interacting model minus the effects of interaction
(to make the continuity equation compatible with redshift-space distortion
measurements). This represents contributions from two averages of the two
coupling constants $\xi_c$ and $\xi_d$; one is weighted by the background
densities of \gls{de} and \gls{dm}, the other, with opposite sign, weighted by
the perturbation to the densities.

Keeping the assumption that galaxies trace the matter field according to $\delta_g = b \, \delta_m$ and $\theta_g = \theta_m = \theta$, the galaxy continuity equation is now 
\begin{align} 
    \mathcal{H} \, \tilde{\beta} \, \delta_g + \theta = 0,
\end{align} 
where $\tilde{\beta} \equiv \tilde{f}_{m} /b$ is the quantity that must replace $\beta$ in eq.~(\ref{eq:powerspectrum}) for the interacting model.

\section{The Fisher matrix formalism} 
\label{sec:Fisher}
The Fisher matrix for the parameters $\vartheta_i$ of a model $\mathcal{M}$ is defined as the ensemble average of the Hessian matrix of the log likelihood. Assuming Gaussian fields with zero mean and covariance {\bf C}, the Fisher matrix is given by
\begin{align}
F_{ij} \equiv \left\langle-\frac{\partial\ln \mathcal{L}}{\partial\vartheta_i \, \partial\vartheta_j}\right\rangle = \frac{1}{2}\mathrm{Tr}\left[\mathrm{\bf C}^{-1}\frac{\partial\mathrm{\bf C}}{\partial\vartheta_i}\mathrm{\bf C}^{-1}\frac{\partial\mathrm{\bf C}}{\partial\vartheta_j}\right] .
\end{align}
For the case of a galaxy power spectrum, the Fisher matrix components for a single tracer results in
\begin{align}
\label{eq:fisher_galaxy} 
	F_{g,\,ij}
    = \frac{1}{2} \int \frac{\ud^3 k}{(2\uppi)^3} \, V_{g,\,\rmn{eff}} \,
    \frac{\partial  \ln P_{g,\,\rmn{obs}}}{\partial \vartheta_i} \frac{\partial
        \ln P_{g,\,\rmn{obs}}}{\partial \vartheta_j}  .
\end{align}
The effective volume is defined as \citep{feldman_power-spectrum_1994,tegmark_measuring_1997}
\begin{align} 
\label{eq:v_eff2} 
    V_{g,\,\rmn{eff}} (k) = \int_V \ud^3 x
    \, \left[ \frac{n_{g}(z)P_{g}(k,z)}{1 + n_{g}(z)
            P_{g}(k,z)} \right]^2,
\end{align} 
where $n_{g}(z)$ is the number density of galaxies and $P_{g}(k,z)$ is the galaxy power spectrum given by eq.~(\ref{eq:powerspectrum}).
However, in practice the information that can be extracted from photometric surveys is limited both by the photometric redshift accuracy and the mode-mixing that takes place due to non-linear structure formation, and for these reasons we redefine the effective volume as
\begin{align} 
\label{eq:v_eff3} 
    V_{g,\,\rmn{eff}} \to &\int_V \ud^3 x \left[ \frac{n_{g}(z)P_{g}(k,z)}{1 + n_{g}(z) P_{g}(k,z)} \right]^2  e^{-2\mu^2\delta_z^2k^2\left(\frac{1 + z}{H(z)}\right)^2 }  e^{-k^2 \Sigma_{\bot}^2 - k^2 \hat{\mu}^2 \left(\Sigma_{\parallel}^2-\Sigma_{\bot}^2 \right) } \,.
\end{align} 
The first exponential factor in eq.~(\ref{eq:v_eff3}) comes
from assuming Gaussian errors for the photometric redshifts,
with variance $\sigma_z = \delta_z \left(1+z \right)$.
The second exponential factor yields a cut-off to avoid
non-linear scales \citep{takada_extragalactic_2014}, where
$\Sigma_{\parallel} = c_{\rmn{rec}} D(z) \, \Sigma_{0}$ and
$\Sigma_{\bot} = c_{\rmn{rec}}D(z)\left(1 + f\right)\Sigma_{0}$.
The constant $c_{\rmn{rec}}$ is introduced to model
the reconstruction method of the \gls{bao} peaks. Without reconstruction,
$c_{\rmn{rec}} = 1$, which is the value we assume in this paper. $D(z)$ is the
growth function normalized as $D(z = 0) = 1$ and we use $\Sigma_{0} = 11 \,
h^{-1} \, \si{\mega\parsec}$.

The Fisher matrix given by eqs.~(\ref{eq:fisher_galaxy}) and (\ref{eq:v_eff3})
allows us to define the Fisher information density per unit
of phase space volume $\left( 2\uppi \right)^{-3} \ud^3 x \, \ud^3 k$ as
(aside from the phenomenological exponential factors)
\begin{equation}
\Phi = \frac{1}{2}\left[\frac{n_{g}(z)P_{g}(k,z)}{1 + n_{g}(z)P_{g}(k,z)}\right]^{2} .
\end{equation}
For surveys which are able to combine multiple tracers of large-scale structure, the Fisher information density can be generalized
to \citep{abramo_why_2013}
\begin{equation}
\Phi_{\alpha\beta}(x,k) = \frac{1}{4}\left[\delta_{\alpha\beta}U_\alpha X + U_\alpha U_\beta\left(1 - X\right)\right] ,
\end{equation}
where $\alpha, \beta = 1 , \dots , N$ are the different types of galaxies, $X_\alpha = n_\alpha P_\alpha(k,z)$ such that $X = \sum_\alpha X_\alpha$, and $U_\alpha = X_\alpha/(1 + X)$. Hence, the Fisher matrix can be generalized for a multi-tracer analysis as
\begin{align}
\label{eq:fisher_multi} 
    F_{ij} = &\sum_{\alpha ,\, \beta = 1}^N \int \frac{\ud^3 x \, \ud^3 k}{(2\uppi)^3} \, \frac{\partial  \ln X_\alpha}{\partial \vartheta_i} \Phi_{\alpha\beta} \frac{\partial \ln X_\beta}{\partial \vartheta_j} e^{-\mu^2\delta_{z,\,\alpha}^2 k^2\left(\tfrac{1 + z}{H(z)}\right)^2 } e^{-\mu^2\delta_{z,\,\beta}^2 k^2 \left(\tfrac{1 + z}{H(z)}\right)^2 } \nonumber \\
    {} &\times 
    \exp\left[-k^2 \Sigma_{\bot}^2 - k^2 \hat{\mu}^2 (\Sigma_{\parallel}^2-\Sigma_{\bot}^2) \right] .
\end{align}
Using this expression we can properly take into account multiple tracers in our
analysis. For instance, we can combine the expected results from \gls{lrg},
\gls{elg} and \gls{qso}, all together.

Another important result is that we can transform a Fisher matrix defined in
terms of a set of $N_\vartheta$ parameters, $\{\vartheta^i\}$, into a set of
$N_\varphi$ parameters, $\{\varphi^\alpha\}$, as long as $N_{\varphi} \leqslant
N_{\vartheta}$.
The Fisher matrix transformation is defined by
\begin{equation}
F_{\alpha\beta} = \sum_{i, j}^{N_\vartheta}\frac{\partial \vartheta^i}{\partial \varphi^\alpha}F_{ij}\frac{\partial \vartheta^j}{\partial \varphi^\beta}.
\end{equation}

In our analysis, we begin with a set of parameters $\vartheta^i = \left\lbrace
\ln{H(z)}, \ln{D_A(z)}, f_s(z), \sigma_{8,\,g}(z), P_{\rmn{shot}}, \Omega_b h^2,
\Omega_c h^2, h, n_s \right\rbrace$, where $f_s(z) =
\tilde{f}_m(z)\sigma_{8,\,m}(z)$. Note that some parameters are local, i.e. they
assume different values at each redshift bin, while others are global. For a
multi-species analysis, each tracer has its own bias and, hence, different
values of $\sigma_{8,\,g}$. 
Later, we marginalize over all those parameters except $\ln{H(z)}$,
$\ln{D_A(z)}$ and $f_s(z)$, which carry all the information about \gls{bao} and
\gls{rsd}. Finally, we project from those parameters to our final set of
cosmological parameters, which are given here by $\Omega_d$, $w$, $\xi_c$ and $\xi_d$.

Our fiducial cosmology is a flat \lcdm\ model with a physical baryon density $\Omega_b
h^2 = \num{0.0226}$, cold dark matter density $\Omega_c h^2 = \num{0.121}$, and
neutrino density parameter, $\Omega_{\nu} h^2 = \num{0.00064}$ (assuming only
one massive neutrino). The reduced Hubble constant is $h = \num{0.68}$ (with a
prior $\sigma_h = \num{0.1}$), \gls{de} equation of state $w = -1$, amplitude
parameter $A_s = \num{2.1e-9}$ and scalar spectral index $n_s = \num{0.96}$.
Planck priors are used to calibrate the \gls{bao} scale only.

Our Fisher code receives as input background and perturbed quantities such as the Hubble rate and the linear matter power spectrum, which were calculated using a modified version of the CAMB code \citep{lewis_efficient_2000, costa_observational_2014, costa_testing_2014} that takes into account the necessary modifications for an interacting dark energy model.

\begin{figure*}[H]
    \centering
    \includegraphics[width=0.85\textwidth]{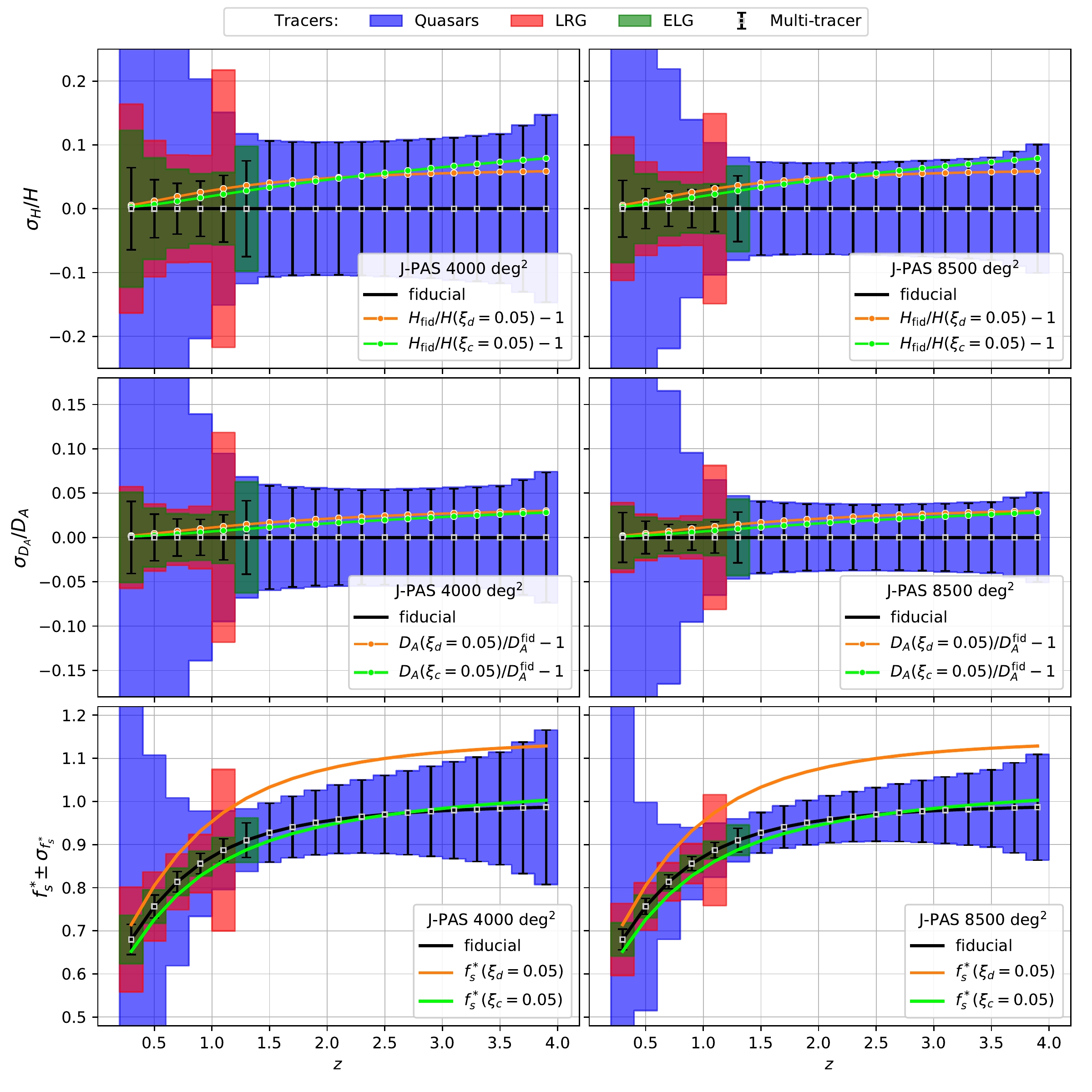}
    \caption{Predicted uncertainties on $H(z)$, $D_A(z)$ and $f_s^*(z) \equiv f_s(z)/{\sigma_{8,m}^{\rmn{fid}}(z)}$ from \gls{lrg}, \gls{elg}, \gls{qso} and the three tracers combined, considering the survey areas of \SIlist{4000;8500}{\square\deg}. We also show the effect, on those functions, of an interaction strength of magnitude \num{0.05} in the cases $Q \propto \rho_d$ and $Q \propto \rho_c$.}
    \label{fig:HDabeta_M2}
\end{figure*}

\section{Results}
\label{sec:results}

We now present the expected constraints on the parameters, as well as the
two-parameter joint constraints for the different cases of our interacting
model. Two scenarios are considered: one using information from $H(z)$ and $D_A(z)$ only, and the other adding information from $f_s(z)$ besides
$H(z)$ and $D_A(z)$. The two cases are labelled ``without RSD'' and ``with RSD'' and, in both, we run our analysis for the two J-PAS survey areas described in section~\ref{sec:data}. We are also conservative
regarding the number density of \glsentrylong{qso}, taking 90 per cent of the densities predicted in previous work
\citep{abramo_measuring_2012}, and the ability to
recover quasi-non-linear scales (reconstruction), which means that if we achieve successful reconstruction with \gls{jpas}, then the constraints from the actual data could be further improved.

Note that we are more concerned here with the marginalized uncertainties on the parameters, under the assumption that they should not vary considerably over the parameter space, i.e., they are not strongly dependent on the choice of fiducial parameters. In fact, one should note that these results do not include any prior information about the allowed region for $\Omega_d$, $w$ and $\xi_{i}$, which will certainly not be true in the actual data analysis when we will have to restrict the parameters to the stability regions listed in Table~\ref{tab:stabilities}.

The uncertainties on $H(z)$, $D_A(z)$ and $f_s(z)$ are shown in Fig.~\ref{fig:HDabeta_M2} together with the effect of the interaction on these functions. We can see that an interacting dark energy induces deviations from our fiducial cosmology. This effect is stronger for higher redshifts in $H(z)$, $D_A(z)$ and $f_s(z)$ with $Q \propto \rho_d$. Thus, \glsentrylong{qso} at high redshifts are expected to produce competitive constraints on the interaction in our models.

Before we discuss our results, we would like to emphasize
that when the interaction is proportional to the \gls{de}
density, we have two distinct regions of stability. One is
characterized by the \gls{de} equation of state in the
phantom regime $w < -1$, in which case the interaction must
be positive, and the other quintessence-like case $-1 < w <
0$, for which $Q$ must be negative. In the Fisher matrix analysis we perform here, there is no need to make explicit those two regions separately. The results
are consistent with one another and, hence, we will not make
such distinction hereafter. The reader, however, must be aware of the stable regions acoording to Table~\ref{tab:stabilities}. For all the results below we use
conservative Gaussian priors on the uncertainties of
$\Omega_d$, $w$ and $\xi$, with variances $\sigma_{\Omega_d}
= 1$, $\sigma_w = 3$ and $\sigma_{\xi} = 1$.  

\begin{figure}
    \centering
    \includegraphics[width=0.95\columnwidth]{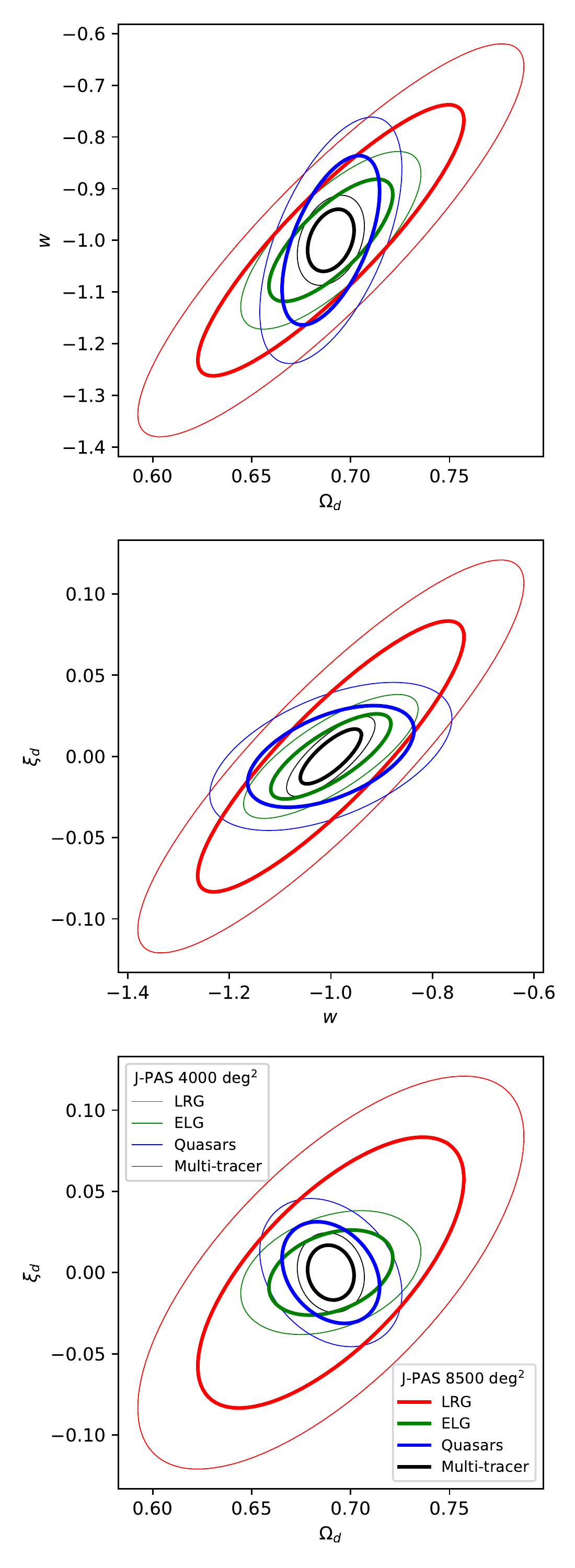}
    \caption{Case $Q \propto \rho_d$ ($\xi \equiv \xi_d$). The ellipses
        represent the 68 per cent uncertainty around the fiducial
        \lcdm\ model. The thin and thick lines correspond to results
        considering the survey areas of \SIlist{4000;8500}{\square\deg},
        respectively. The red contours are for \gls{lrg}, green for \gls{elg},
        blue for \gls{qso}, and black for the Multi-tracer analysis,
        all with
        \gls{rsd}.}
    \label{fig:M1joint}
\end{figure}

\begin{table}
    \renewcommand*{\arraystretch}{1.2}
    \centering
    \caption{Marginalized uncertainties for the three surveys, without
        \gls{rsd}, for the case where the interacting coupling term is proportional to dark energy density, $Q \propto \rho_d$. The parameters we are concerned, $\Omega_d$, $w$ and $\xi_d$  are obeserved to be very degenerated. The results are dominated by our priors.}
    \label{tab:margerrors_model1_noRSD}
    \sisetup{
        round-mode=places,
        round-precision=3,
        table-format=1.3
    }
    \begin{tabular}{@{}lcSSSS@{}}
        \hline
        &   Uncertainty   &   {\gls{lrg}} & {\gls{elg}} & {\gls{qso}} & {Multi-tracer} \\
        \hline
        \multirow{3}{*}{\begin{minipage}[t]{0.45 in} \gls{jpas} \\ {\footnotesize (\SI{4000}{\square\deg})} \end{minipage}} %
        &   $\sigma_{\Omega_d}$  &   0.551611   &   0.547215    &    0.546236    &    0.546139    \\
        &   $\sigma_w$  &   0.855581    &    0.813336    &    0.811246    &    0.798096    \\
        &   $\sigma_{\xi_d}$    &   0.799965    &    0.796182    &    0.795041    &    0.794886    \\
        \hline
        \multirow{3}{*}{\begin{minipage}[t]{0.45 in} \gls{jpas} \\ {\footnotesize (\SI{8500}{\square\deg})} \end{minipage}} %
        & $\sigma_{\Omega_d}$ &   0.548625    &    0.546538    &    0.546164    &    0.546074    \\
        &  $\sigma_{w}$    &   0.823799    &    0.803137    &    0.802853    &    0.796178    \\
        &  $\sigma_{\xi_d}$    &   0.797374    &    0.795525    &    0.794837    &    0.794794    \\
        \hline
        \multirow{3}{*}{DESI}  &   $\sigma_{\Omega_d}$ &   0.54692 &   0.546083 &   0.546359 &   0.54604 \\
        &   $\sigma_{w}$    &   0.808902 &   0.79697 &   0.812673 &   0.795835   \\
        &   $\sigma_{\xi_d}$    &   0.795916     &   0.79498     &   0.79539     &   0.794928     \\
        \hline
        \multirow{3}{*}{Euclid}  &   $\sigma_{\Omega_d}$ &       &   0.546013    &   &       \\
        &   $\sigma_{w}$    &       &   0.795061    &   &      \\
        &   $\sigma_{\xi_d}$    &       &    0.794784   &  &      \\
        \hline
    \end{tabular}
\end{table}
\begin{table}
    \renewcommand*{\arraystretch}{1.2}
    \centering
    \caption{Marginalized uncertainties for the three surveys, with \gls{rsd}, for the case where the interacting coupling term is proportional to dark energy density, $Q \propto \rho_d$. The inclusion of RSD information breaks the strong degeneracy presented before.}
    \label{tab:margerrors_model1_RSD}
    \sisetup{
        round-mode=places,
        round-precision=3,
        table-format=1.3
    }
    \begin{tabular}{@{}lcSSSS@{}}
        \hline
        &   {Uncertainty}   &   {\gls{lrg}} & {\gls{elg}} & {\gls{qso}} & {Multi-tracer} \\
        \hline
        \multirow{3}{*}{\begin{minipage}[t]{0.45 in} \gls{jpas} \\ {\footnotesize (\SI{4000}{\square\deg})} \end{minipage}} %
        &   $\sigma_{\Omega_d}$  &   0.0643925   &   0.0300546   &   0.0237013    &   0.0112807  \\
        &   $\sigma_w$  &   0.251013    &   0.113509   &   0.157433    &   0.0577221   \\
        &   $\sigma_{\xi_d}$    &   0.0798392   &   0.025182   &   0.0300985   &   0.0162501   \\
        \hline
        \multirow{3}{*}{\begin{minipage}[t]{0.45 in} \gls{jpas} \\ {\footnotesize (\SI{8500}{\square\deg})} \end{minipage}} %
        & $\sigma_{\Omega_d}$ &   0.0443724   &   0.0206666   &   0.0162713 &   0.00775209   \\
        &  $\sigma_{w}$    &   0.173152    &   0.0781386   &   0.108152 &   0.0397262   \\
        &  $\sigma_{\xi_d}$    &   0.0550296   &   0.0173067   &   0.0206597 &   0.0111527 \\
        \hline
        \multirow{3}{*}{DESI}  &   $\sigma_{\Omega_d}$ &   0.0318607   &   0.0125006   &   0.0262382   &   0.0104975   \\
        &   $\sigma_{w}$    &   0.14543    &   0.0579946   &   0.161782    &   0.0467318   \\
        &   $\sigma_{\xi_d}$    &   0.0411012   &   0.00911882  &   0.0258308   &   0.0077904  \\
        \hline
        \multirow{3}{*}{Euclid}  &   $\sigma_{\Omega_d}$ &       &   0.00608301    &       &     \\
        &   $\sigma_{w}$    &       &   0.0280099    &       &      \\
        &   $\sigma_{\xi_d}$    &       &    0.00419022    &       &      \\
        \hline
    \end{tabular}
\end{table}
The marginalized constraints for the case $Q \propto \rho_d$ are shown in Tables~\ref{tab:margerrors_model1_noRSD} (without \gls{rsd}) and \ref{tab:margerrors_model1_RSD} (with \gls{rsd}). We present the results for two
\gls{jpas} areas, together with the expected results for DESI and Euclid.
We observe that, when information from \gls{rsd} is not considered, our three
parameters of interest are very degenerate.
The constraints are dominated by our priors on $\Omega_d$, $w$ and $\xi_d$.
None of the tracers nor any survey was able to
break this degeneracy and produce significant constraints.
However, the inclusion of \gls{rsd} introduces new information that alleviates
the degeneracy.
In this case, the prior uncertainties are not important and we obtain constraints
of a few per cent as observed in Table~\ref{tab:margerrors_model1_RSD}. This can be compared with a similar analysis made in \cite{Santos:2017bqm} for an Euclid-like survey. We can see that our constraints for the DE equation of state and the interaction parameter are comparable to those found in their Tables III and IV. Comparing the multi-tracer analysis for the
three surveys (actually we are not using multi-tracer for Euclid, only ELG), we see that \gls{jpas} can produce slightly better constraints than DESI for the dark energy density parameter and the equation of state. This is associated with the expected values for \gls{qso}, which are denser and reach higher redshifts in \gls{jpas}. The joint constraints for $\Omega_d$, $w$ and $\xi_d$ are shown in Fig.~\ref{fig:M1joint} for the two areas and for the different tracers with \gls{jpas}.

\begin{figure}
    \centering
    \includegraphics[width=0.95\columnwidth]{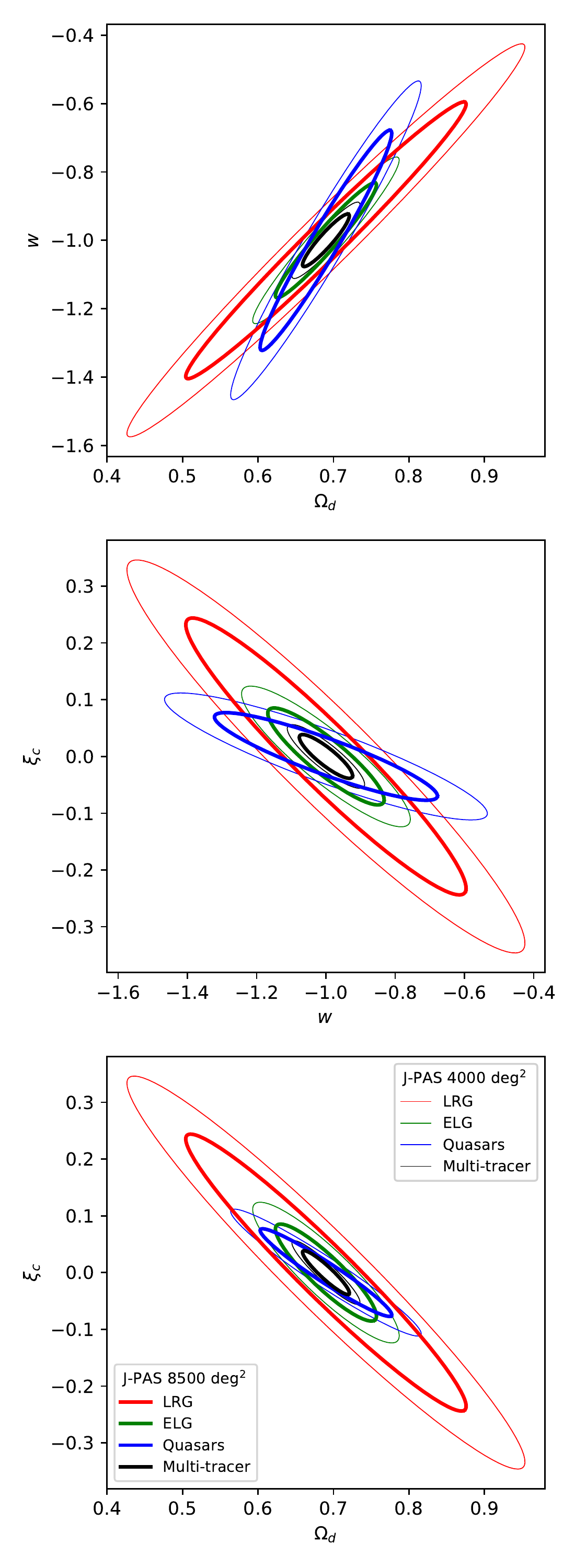}
    \caption{Case $Q \propto \rho_c$ ($\xi \equiv \xi_c$). The ellipses
        represent the 68 per cent uncertainty around the fiducial
        \lcdm\ model. The thin and thick lines correspond to results
        considering the survey areas \SIlist{4000;8500}{\square\deg},
        respectively. The red contours are for \gls{lrg}, green for \gls{elg},
        blue for \gls{qso}, and black for the Multi-tracer analysis,
        all with \gls{rsd}.}
    \label{fig:M3joint}
\end{figure}

\begin{table}
    \renewcommand*{\arraystretch}{1.2}
    \centering
    \caption{Marginalized uncertainties for the three surveys, without \gls{rsd}, for the case where the interacting coupling term is proportional to dark matter density, $Q \propto \rho_c$. J-PAS (\SI{8500}{\square\deg}) can put better constraints than DESI and Euclid in the multi-tracer analysis.}
    \label{tab:margerrors_model3_noRSD}
    \sisetup{
        round-mode=places,
        round-precision=3,
        table-format=1.3
    }
    \begin{tabular}{@{}lcSSSS@{}}
        \hline
        &   {Uncertainty}   &   {\gls{lrg}} & {\gls{elg}} & {\gls{qso}} & {Multi-tracer} \\
        \hline
        \multirow{3}{*}{\begin{minipage}[t]{0.45 in} \gls{jpas} \\ {\footnotesize (\SI{4000}{\square\deg})} \end{minipage}} %
        &  $\sigma_{\Omega_d}$ &   0.453118 &   0.417929    &   0.218353 &   0.115342   \\
        &   $\sigma_{w}$    &   0.844057 &   0.787531 &   0.620995    &   0.272808    \\
        &   $\sigma_{\xi_c}$    &   0.830275     &   0.711576    &   0.230388    &   0.135405    \\
        \hline
        \multirow{3}{*}{\begin{minipage}[t]{0.45 in} \gls{jpas} \\ {\footnotesize (\SI{8500}{\square\deg})} \end{minipage}} %
        &  $\sigma_{\Omega_d}$ &   0.435129    &   0.367391 &   0.155782    &   0.0799841   \\
        &   $\sigma_{w}$    &   0.788325 &   0.687092 &   0.442607 &   0.189143    \\
        &   $\sigma_{\xi_c}$    &   0.804001 &   0.626271    &   0.164268    &   0.0938781   \\
        \hline
        \multirow{3}{*}{DESI} &   $\sigma_{\Omega_d}$ &   0.419272    &   0.215231 &   0.360715    &   0.143934 \\
        &   $\sigma_w$  &   0.775474 &   0.435715    &   0.80686 &   0.291974    \\
        &   $\sigma_{\xi_c}$    &   0.746112     &   0.332556     &   0.499914    &   0.220489    \\
        \hline
        \multirow{3}{*}{Euclid} &   $\sigma_{\Omega_d}$ &       &   0.0962103    &    &       \\
        &   $\sigma_w$  &       &   0.209334    &     &     \\
        &   $\sigma_{\xi_c}$    &       &       0.133743   &     &     \\
        \hline
    \end{tabular}
\end{table}
\begin{table}
    \renewcommand*{\arraystretch}{1.2}
    \centering
    \caption{Marginalized uncertainties for the three surveys, with \gls{rsd}, for the case where the interacting coupling term is proportional to dark matter density, $Q \propto \rho_c$.}
    \label{tab:margerrors_model3_RSD}
    \sisetup{
        round-mode=places,
        round-precision=3,
        table-format=1.3
    }
    \begin{tabular}{@{}lcSSSS@{}}
        \hline
        &   {Uncertainty}   &   {\gls{lrg}} & {\gls{elg}} & {\gls{qso}} & {Multi-tracer} \\
        \hline
        \multirow{3}{*}{\begin{minipage}[t]{0.45 in} \gls{jpas} \\ {\footnotesize (\SI{4000}{\square\deg})} \end{minipage}} %
        &  $\sigma_{\Omega_d}$ &   0.174027    &   0.064073    &   0.0833263   &   0.0298323   \\
        &   $\sigma_{w}$    &   0.379296    &   0.160843    &   0.307698    &   0.0739509   \\
        &   $\sigma_{\xi_c}$    &   0.228477    &   0.0818084   &   0.0738403   &   0.0373186    \\
        \hline
        \multirow{3}{*}{\begin{minipage}[t]{0.45 in} \gls{jpas} \\ {\footnotesize (\SI{8500}{\square\deg})} \end{minipage}} %
        &  $\sigma_{\Omega_d}$ &   0.122625   &   0.0441925   &   0.0575563   &   0.0205179   \\
        &   $\sigma_{w}$    &   0.267129    &   0.110926   &   0.212541    &   0.0508994   \\
        &   $\sigma_{\xi_c}$    &   0.160991    &   0.0563771   &   0.0509725   &   0.0256439   \\
        \hline
        \multirow{3}{*}{DESI} &   $\sigma_{\Omega_d}$ &   0.122432    &   0.0297094    &   0.0755915   &   0.0250848   \\
        &   $\sigma_w$  &   0.284144 &   0.086188   &   0.243457    &   0.0714552   \\
        &   $\sigma_{\xi_c}$    &   0.145539    &   0.030383   &   0.0763357   &   0.02613   \\
        \hline
        \multirow{3}{*}{Euclid} &   $\sigma_{\Omega_d}$ &       &   0.0124031    &    &       \\
        &   $\sigma_w$  &       &       0.0394763    &       &     \\
        &   $\sigma_{\xi_c}$    &       &    0.0132333     &     &     \\
        \hline
    \end{tabular}
\end{table}
The same is done for the case $Q \propto \rho_c$, presented in Tables~\ref{tab:margerrors_model3_noRSD} (without \gls{rsd}) and \ref{tab:margerrors_model3_RSD} (with \gls{rsd}) and in Fig.~\ref{fig:M3joint}. In this case, the resulting parameters are not as degenerate as in the previous case. The
constraints on $H(z)$ and $D_A(z)$ can provide significant information and our
prior is not as dominant as before. For instance, in the multi-tracer analysis,
the prior uncertainties only alter our results at $\sim 2$ per cent for
\gls{jpas} $\SI{4000}{\square\deg}$ and at $\sim 1$ per cent for \gls{jpas}
$\SI{8500}{\square\deg}$. Table 7 shows us that J-PAS can put better constraints than DESI and Euclid when we only consider BAO information in this model. Again, the constraints from LRG, ELG and QSO indicate that QSO are playing the role in this leadership here. Including information from \gls{rsd} can improve the
results even more. However, the constraints in this case are not as sensitive to \gls{rsd} as when $Q \propto \rho_d$, as could be expected from Fig.~\ref{fig:HDabeta_M2}. J-PAS still provides better results than DESI with RSD information, but Euclid gives an uncertainty on the coupling constant twice as better. Also, in this case, we can compare our results with \citet{Santos:2017bqm}. This time, our constraints in Table~\ref{tab:margerrors_model3_RSD} are weaker than those found by them in their Table V. In fact, their constraint for the interaction parameter is three times stronger than ours. However, this must be related with some differences between our analysis, as some dependence with the fiducial values. In any case, we have obtained a more conservative result.

\begin{table}
    \setlength\tabcolsep{6pt}
    \renewcommand*{\arraystretch}{1.2}
    \caption{Marginalized uncertainties for the three surveys, without
        \gls{rsd}, for the case where the interacting coupling term is proportional to the sum of the dark sector energy densities, $Q \propto \rho_c + \rho_d$ ($\xi_c = \xi_d \equiv
        \xi$).}
    \label{tab:margerrors_model4_noRSD}
    \centering
    \sisetup{
        round-mode=places,
        round-precision=3,
        table-format=1.3
    }
    \begin{tabular}{@{}lcSSSS@{}}
        \hline
        &   {Uncertainty}   &   {\gls{lrg}} & {\gls{elg}} & {\gls{qso}} & {Multi-tracer} \\
        \hline
        \multirow{3}{*}{\begin{minipage}[t]{0.45 in} \gls{jpas} \\ {\footnotesize (\SI{4000}{\square\deg})} \end{minipage}} %
        &   $\sigma_{\Omega_d}$  &   0.707856 &   0.670141 &   0.353694    &   0.203799    \\  
        &   $\sigma_{w}$    &   1.14623 &   1.10284 &   0.794619    &   0.395584    \\  
        &   $\sigma_{\xi}$  &   0.58476 &   0.528425 &   0.216951    &   0.132972    \\
        \hline
        \multirow{3}{*}{\begin{minipage}[t]{0.45 in} \gls{jpas} \\ {\footnotesize (\SI{8500}{\square\deg})} \end{minipage}} %
        &    $\sigma_{\Omega_d}$ &   0.697195   &   0.6227  & 0.259794    &   0.142753 \\
        &   $\sigma_{w}$    &   1.11693 &   1.02155 &   0.583123    &   0.27704    \\
        &   $\sigma_{\xi}$  &   0.572847     &   0.490286    &   0.15925    &   0.0931193   \\
        \hline
        \multirow{3}{*}{DESI} &   $\sigma_{\Omega_d}$ &   0.67887 & 0.406324    & 0.578443    & 0.283681    \\
        &   $\sigma_{w}$    &   1.1041 & 0.70097    & 1.0689 & 0.489726    \\
        &   $\sigma_{\xi}$  &   0.545598 & 0.304971    & 0.411653    & 0.212106    \\
        \hline
        \multirow{3}{*}{Euclid} &   $\sigma_{\Omega_d}$  &       &   0.185087    &     &     \\
        &   $\sigma_{w}$    &       &   0.336319    &     &      \\
        &   $\sigma_{\xi}$  &       &   0.131775    &     &    \\
        \hline
    \end{tabular}
\end{table}
\begin{table}
    \setlength\tabcolsep{6pt}
    \renewcommand*{\arraystretch}{1.2}
    \caption{Marginalized uncertainties for the three surveys, with
        \gls{rsd}, for the case where the interacting coupling term is proportional to the sum of the dark sector energy densities, $Q \propto \rho_c + \rho_d$ ($\xi_c = \xi_d \equiv
        \xi$).}
    \label{tab:margerrors_model4_RSD}
    \centering
    \sisetup{
        round-mode=places,
        round-precision=3,
        table-format=1.3
    }
    \begin{tabular}{@{}lcSSSS@{}}
        \hline
        &   {Uncertainty}   &   {\gls{lrg}} & {\gls{elg}} & {\gls{qso}} & {Multi-tracer} \\
        \hline
        \multirow{3}{*}{\begin{minipage}[t]{0.45 in} \gls{jpas} \\ {\footnotesize (\SI{4000}{\square\deg})} \end{minipage}} %
        &   $\sigma_{\Omega_d}$  &   0.050504   &   0.0301982   &   0.0575123   &   0.0188978  \\  
        &   $\sigma_{w}$    &   0.17809    &   0.0944343   &   0.141167    &   0.0410643  \\  
        &   $\sigma_{\xi}$  &   0.110027   &   0.0349267 &   0.0405026   &   0.0187075  \\
        \hline
        \multirow{3}{*}{\begin{minipage}[t]{0.45 in} \gls{jpas} \\ {\footnotesize (\SI{8500}{\square\deg})} \end{minipage}} %
        &    $\sigma_{\Omega_d}$ &   0.0346917   &   0.0207315   &     0.0395224   &   0.0129814   \\
        &   $\sigma_{w}$    &   0.122716   &   0.0650148   &   0.0970109   &   0.028301   \\
        &   $\sigma_{\xi}$  &   0.07584   &   0.0240049   &   0.0278224   &   0.0128428   \\
        \hline
        \multirow{3}{*}{DESI} &   $\sigma_{\Omega_d}$ &   0.0235797   & 0.0120747   & 0.0337923   & 0.0109078   \\
        &   $\sigma_{w}$    &   0.0934715   & 0.0476072   & 0.126902    & 0.0383423   \\
        &   $\sigma_{\xi}$  &   0.0539899   & 0.012611   & 0.0361464   & 0.0106951   \\
        \hline
        \multirow{3}{*}{Euclid} &   $\sigma_{\Omega_d}$  &       &   0.00761294    &     &     \\
        &   $\sigma_{w}$    &       &   0.024792     &     &      \\
        &   $\sigma_{\xi}$  &       &   0.00586946     &     &    \\
        \hline
    \end{tabular}
\end{table}
For the last case considered, $Q \propto \rho_d + \rho_c$, we give the results for the marginalized constraints in Tables~\ref{tab:margerrors_model4_noRSD} (without \gls{rsd}) and \ref{tab:margerrors_model4_RSD} (with \gls{rsd}).
As one could expect, in this scenario we see characteristics combined from both
previous cases. The measurements of $H(z)$ and $D_A(z)$ give significant information,
especially at high redshifts, as in $Q \propto \rho_c$. 
The constraints are also very sensitive to \gls{rsd} as in $Q \propto \rho_d$.

\begin{figure}
    \centering
    \includegraphics[width=.98\columnwidth]{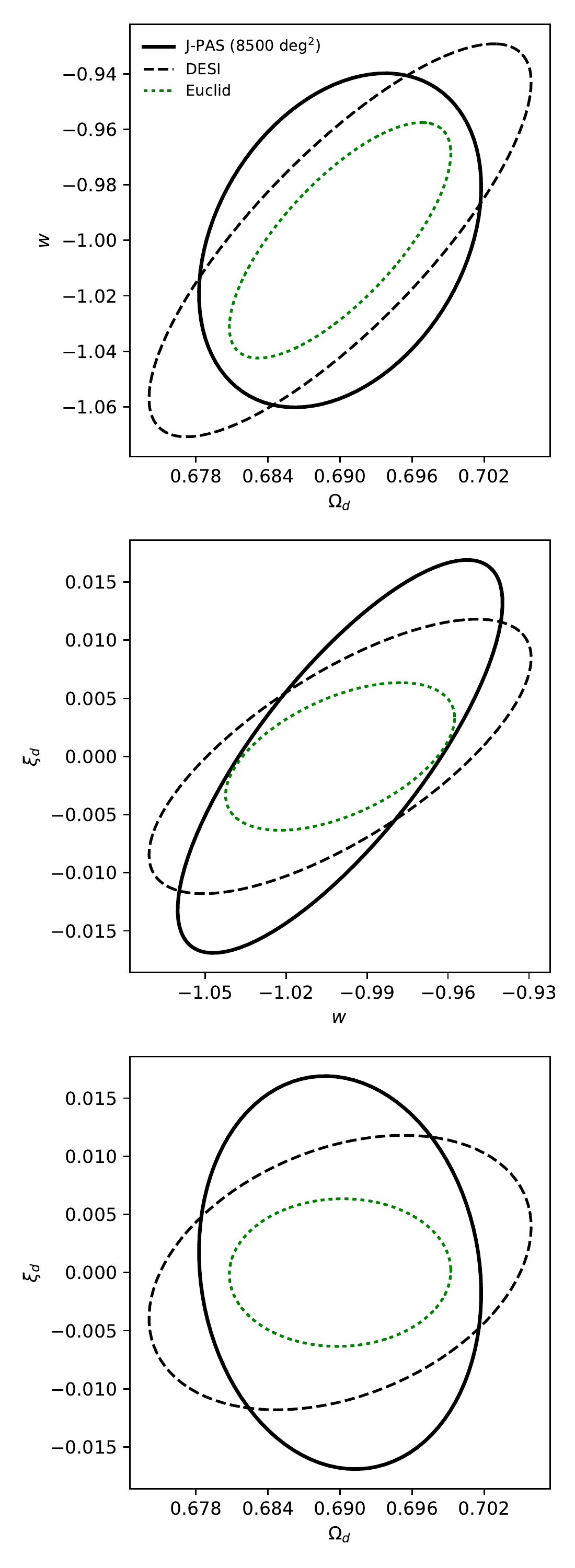}
    \caption{Comparison of the 68 per cent uncertainties around the fiducial \lcdm\ model for the \gls{jpas} (\SI{8500}{\square\deg}), Euclid and DESI surveys including information from BAO and RSD in a multi-tracer analysis. Case $Q \propto \rho_d$ ($\xi \equiv \xi_d$).}
    \label{fig:M1comp}
\end{figure}

\begin{figure}
    \centering
    \includegraphics[width=.98\columnwidth]{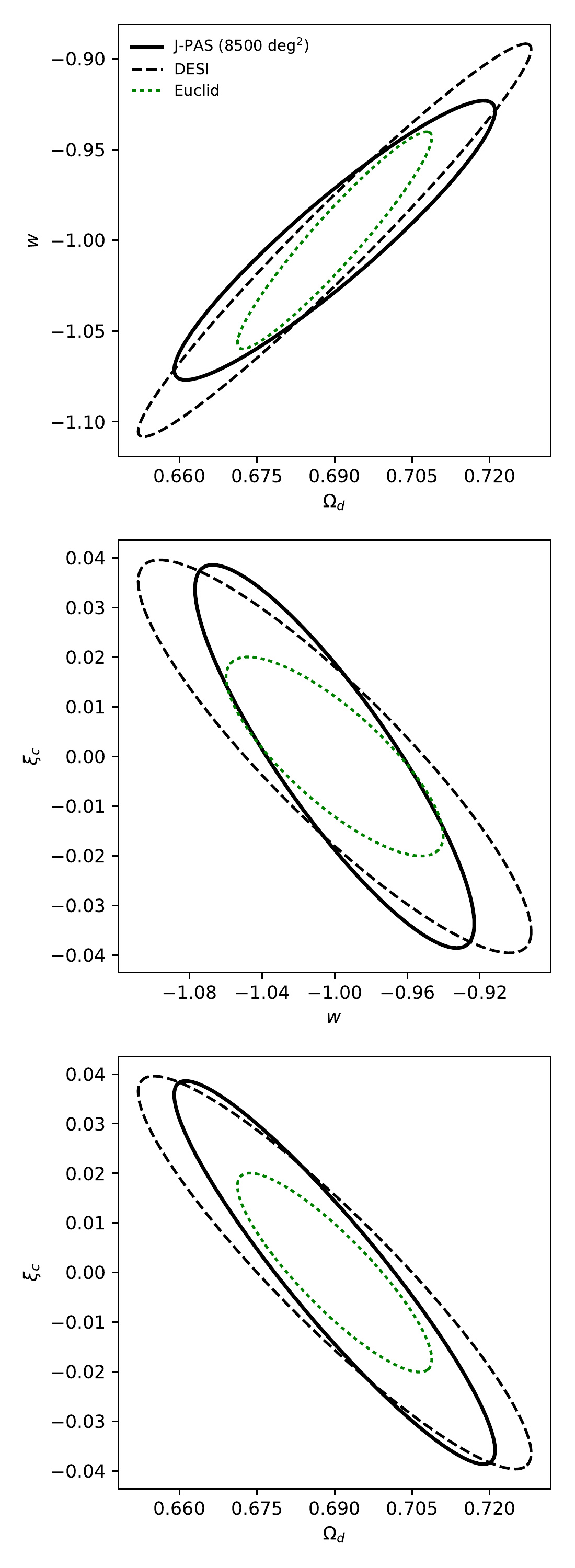}
    \caption{Comparison of the 68 per cent uncertainties around the fiducial \lcdm\ model for the \gls{jpas} (\SI{8500}{\square\deg}), Euclid and DESI surveys including information from BAO and RSD in a multi-tracer analysis. Case $Q \propto \rho_c$ ($\xi \equiv \xi_c$).}
    \label{fig:M3comp}
\end{figure}

Using information from BAO and RSD, we compare the expected confidence regions of \gls{jpas} (\SI{8500}{\square\deg}), DESI and Euclid, all with multiple tracers except Euclid, which has only one kind of tracer. The results for the cases $Q \propto \rho_d$ and $Q \propto \rho_c$ are presented in Figs.~\ref{fig:M1comp} and \ref{fig:M3comp}. Even though Euclid has only one kind of tracer, it shows the best constraints in those figures. This is related to its large survey area, high galaxy number densities and small redshift errors. On the other hand, although DESI covers a larger area in the sky and has a smaller redshift error than J-PAS, their constraints are comparable because of the larger redshift range of J-PAS.
\begin{figure}
    \centering
    \includegraphics[width=\columnwidth]{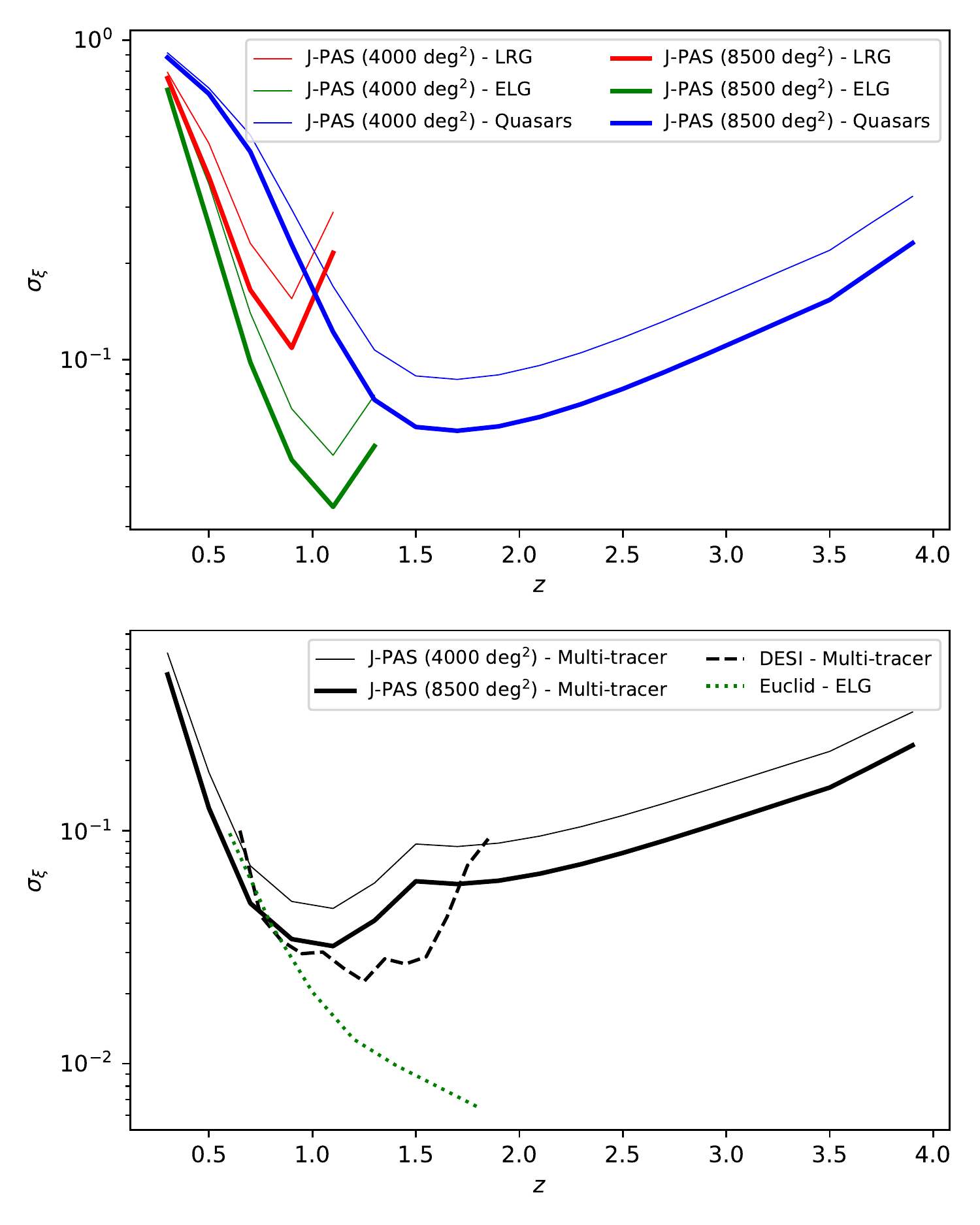}
    \caption{Constraint on $\xi_d$ as a function of the redshift $z$ under the
        different survey configurations,
        case $Q \propto \rho_d$ ($\xi \equiv \xi_d$).}
    \label{fig:M1sigma_z}
\end{figure}
\begin{figure}
    \centering
    \includegraphics[width=\columnwidth]{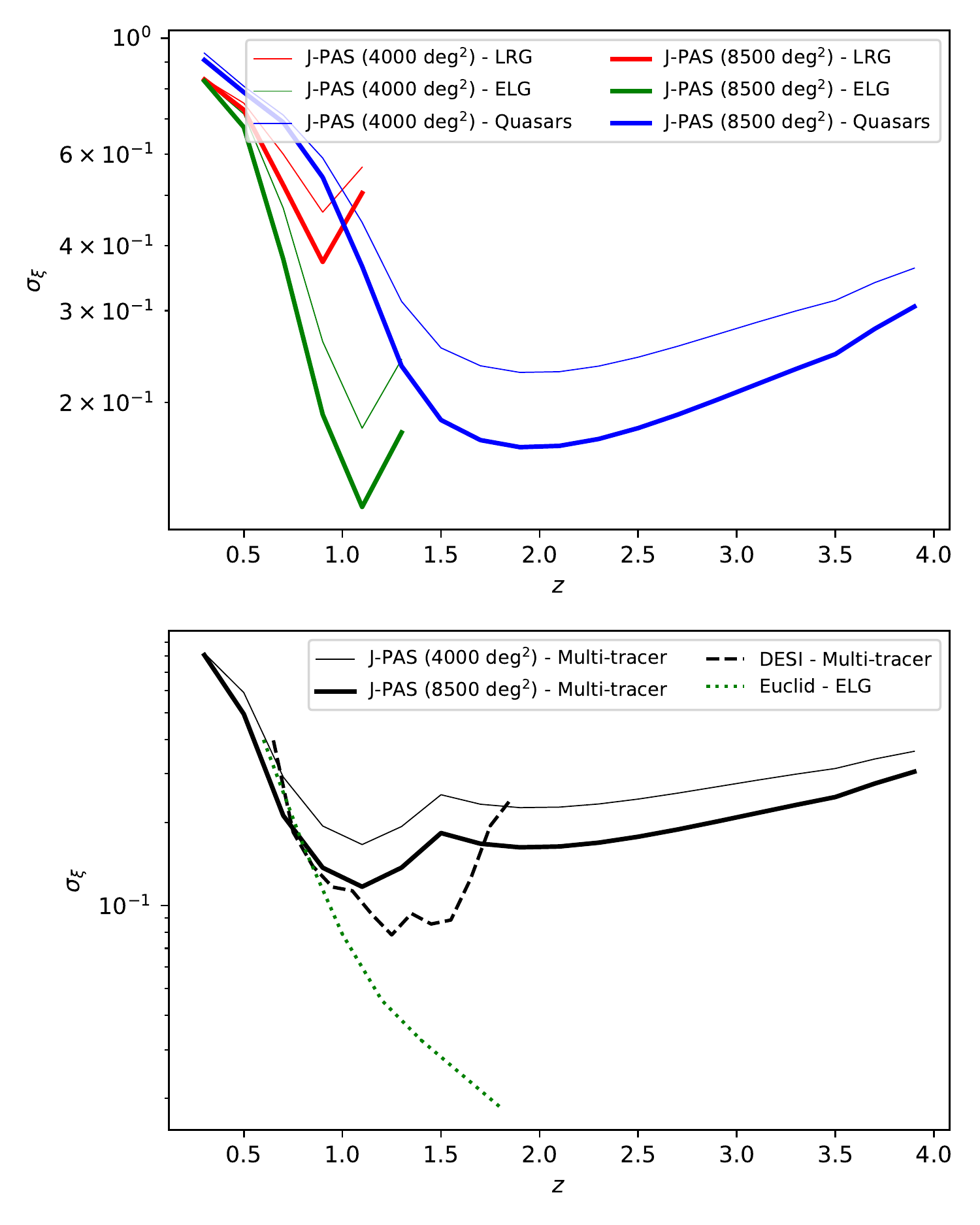}
    \caption{Constraint on $\xi_c$ as a function of the redshfit $z$ under the
        different survey configurations,
        case $Q \propto \rho_c$ ($\xi \equiv \xi_c$).}
    \label{fig:M3sigma_z}
\end{figure}
\begin{figure}
    \centering
    \includegraphics[width=\columnwidth]{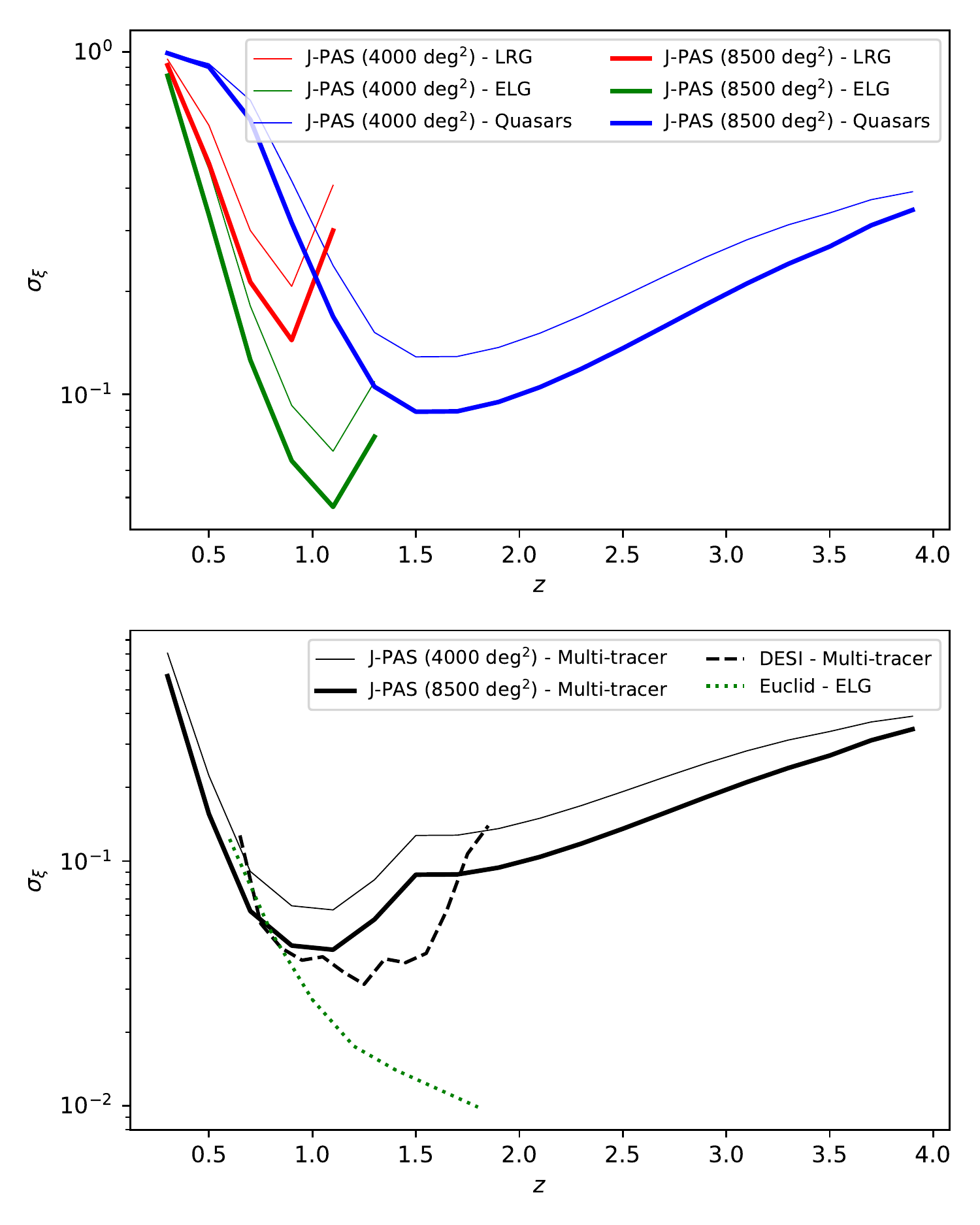}
    \caption{Constraint on $\xi$ as a function of the redshift $z$ under the
        different survey configurations,
        case $Q \propto \rho_c + \rho_d$ ($\xi_c = \xi_d \equiv \xi$).}
    \label{fig:M4sigma_z}
\end{figure}

It is important to notice that, in our analysis, we are not considering the Bright Galaxy Survey (BGS) sample at low redshift from \citet{desi_collaboration_desi_2016}\footnote{We thank Prof. Daniel Eisenstein to call our attention for that fact in a private communication.}. As a simple flux-limited sample of galaxies, it will consist of different types of galaxies. Thus, one of the reasons to avoid this sample in our analysis is how to properly take it into account in our multi-tracer analysis. On the other hand, the basic strategy for galaxy clustering with Euclid does not possess low redshift measurements \citep{laureijs_euclid_2011}, although, it is possible to add low redshift data from other survey, such as the
\glsentrylong{sdss}
\citep[\glsentryshort{sdss},\glsunset{sdss}][]{blanton_sloan_2017}. Therefore, including those low redshift data for DESI and Euclid would improve the constraints found for them in this paper.

Finally, we see that the results for our interacting \gls{de} model are very sensitive to the use of \glsentrylongpl{rsd}. This is certainly true for an interaction proportional to the \gls{de} density, where loose constraints dominated by our priors tighten considerably when we include RSD, but also in the other cases, in which the constraints can improve by a factor of \num{10} in some scenarios. This was also clear in previous publications \citep{murgia_constraints_2016, costa_constraints_2017,li_new_2018}. We note that high redshift measurements tend to place better constraints, as the interaction yields stronger deviations from the standard model at those redshifts. Figs.~\ref{fig:M1sigma_z}, \ref{fig:M3sigma_z} and \ref{fig:M4sigma_z} summarize the interaction constraints as a function of the redshift, for the cases $Q \propto \rho_d$, $Q \propto \rho_c$ and $Q \propto \rho_c + \rho_d$, respectively. However,
combining different tracers at various redshifts in a
multi-tracer analysis have produced the best scenario.
On the other hand, increasing the \gls{jpas} survey area from
\SIrange[range-phrase={ to }]{4000}{8500}{\square\deg}
induces a relative difference on the constraints of about 40 per cent.

\section{Conclusions}
\label{sec:conclusions}
In this work, we use information from \glsentrylongpl{bao} and \glsentrylongpl{rsd} to estimate the constraining power of the J-PAS survey for parameters of an interacting dark energy model. The analysis is done using the Fisher matrix formalism and Planck priors were only used to calibrate the \gls{bao} scale.

Employing the whole galaxy power spectrum, we marginalize over several cosmological
parameters ending up with three local parameters, $\ln{H(z)}$, $\ln{D_A(z)}$ and
$f_s(z)$, which basically carry information about the \gls{bao} scale and
\gls{rsd}. Then, we project the expected constraints on those parameters on
constraints over our interacting dark energy model, which is described by the dark energy density fraction $\Omega_d$, the equation of state
$w$, and the interaction parameter $\xi_c$ or $\xi_d$.

We consider the effect of different tracers (i.e LRG, ELG and QSO) of the underlying matter distribution on the constraints and, also, a multi-tracer analysis. The impact of the survey area is also take into account and the results are compared with those from DESI and Euclid.

We find that, with J-PAS data in the near future, we shall be able to determine the interaction parameter with a maximum precision of $\sigma_{\xi_c} \sim 0.02$ when the interaction term is proportional to the \gls{dm} energy density and of $\sigma_{\xi_d} \sim 0.01$ when the interaction is proportional to the \gls{de} density. These numbers are similar to the constraints predicted by DESI. For the constant equation of state of dark energy, the best predicted constraints from \gls{jpas} are slightly better than those from DESI in both interacting cases: $\sigma_{w}$ about \numrange{0.04}{0.05} against \numrange{0.05}{0.07} around the fiducial value $w = -1$. In terms of constraining power and in the context of our interacting model, both surveys are behind Euclid but get close to it, projecting comparable constraints on the relevant parameters in all specific cases considered.

Finally, we would like to enphasize some limitations and possible extensions of this work:
\begin{itemize}
\item As it is well known, the Fisher matrix formalism provides the best case scenario for a forecast. A natural extension should properly explore the space of parameters as in a Monte Carlo approach. In this case, the unstable regions presented in Table~\ref{tab:stabilities} would be avoided by some priors.

\item Also two aspects that could impair the J-PAS constraints in comparison to DESI and Euclid are a more realistic photo-$z$ error distribution (with longer tails and more outliers than a Gaussian distribution) and the contamination of our galaxy sample by stars (and by tracers of a different type). This will become clearer in the next months with ongoing J-PAS proof-of-concept tests.

\item We have only taken into account contributions from BAO and RSD. However, J-PAS is able to do more. A more complete analysis could combine information from supernovae type Ia, weak lensing and galaxy clusters.

\item At $z \geq 2$, J-PAS will be able to detect a significant population of Lyman $\alpha$ Emmiters (LAEs) (more numerous than QSOs) that is not taken into account in this analysis. This could significantly enhance the importance of high-z constraints.

\item The likelihood function for every survey will depend strongly on the range of scales that is used to measure $P(k)$. This is especially important for RSD analysis. Also, the assembly bias, the description of non-linear density and velocity field regimes, and the impact of galaxy formation in general could make the modeling of RSDs significantly more challenging, e.g. \citet{Orsi:2017ggf}. This could either bias the constraints, or dramatically weaken the contribution of RSDs to the overall constraints.
\end{itemize}

\section*{Acknowledgements} 
A.A.C. acknowledges FAPESP and CAPES for the financial support under grant number 2013/26496-2 (FAPESP). R.G.L.~is supported by CNPq under the grant 208206/2017-5. E. A. acknowledges FAPESP (grant number 2014/07885-0) and CNPq for support. L.R.A. thanks FAPESP and CNPq for financial support. R.A.D. acknowledges support from the Conselho Nacional de Desenvolvimento Cient\'ifico e Tecnol\'ogico - CNPq through BP grant 312307/2015-2, and the Financiadora de Estudos e Projetos - FINEP grants REF. 1217/13 - 01.13.0279.00 and REF 0859/10 - 01.10.0663.00 and also FAPERJ PRONEX grant E-26/110.566/2010 for hardware funding support for the J-PAS project through the National Observatory of Brazil and Centro Brasileiro de Pesquisas F\'isicas.

This paper has gone through internal review by the J-PAS
collaboration. Funding for the J-PAS Project has been
provided by the Governments of Espa\~na and Arag\'on through
the Fondo de Inversi\'on de Teruel, European FEDER funding
and the MINECO and by the Brazilian agencies FINEP, FAPESP,
FAPERJ and by the National Observatory of Brazil.

\bibliographystyle{mnras}
\bibliography{J-PAS-IDE}

\bsp	
\label{lastpage}
\end{document}